\documentstyle[12pt,cite]{article}                                        
\catcode`\@=11                                                       
\def\ee{\end{equation}}
\def\be{\begin{equation}}
\def\la{\label}

\def\H{{\cal H}}

\def\N{{\cal N}}
\def\1{{\bf 1}_N}
\def\End{{\rm End}}

\begin{document}                                                     

\renewcommand{\theequation}{\thesection.\arabic{equation}}        
\newcommand{\mysection}[1]{\setcounter{equation}{0}\section{#1}}  

\begin{titlepage}


\hfill{preprint SPIN -- 2000/33}

\vspace{2 cm}

\centerline{{\huge{The geometry of M5-branes and TQFTs}}}

\vspace{2 cm}

\centerline{\large{Giulio Bonelli}}
\centerline{Spinoza Institute, University of Utrecht}
\centerline{Leuvenlaan 4, 3584, CE Utrecht, The Netherlands}
\centerline{G.Bonelli@phys.uu.nl}

\vspace{4 cm}

{\it Abstract}:
The calculation of the partition function for N M5-branes is addressed 
for the case in which the worldvolume wraps a manifold $T^2\times M_4$,
where $M_4$ is simply connected and Kaehler. This is done in a 
compactification 
of M-theory which induces the Vafa-Witten theory on $M_4$ in the 
limit of vanishing torus volume. 
The results follow from the equivalence of the BPS spectrum counting in 
the complementary limit of vanishing $M_4$ volumes and
from a classification of the the moduli space of quantum 
vacua of the supersymmetric twisted theory in terms of associated spectral covers.
This reduces the problem of the moduli counting to algebraic equations.

\end{titlepage}

\section{Introduction}

The discovery of
D-branes \cite{pol} led string theory to new 
and unexpected roads. 
While the low energy effective theory of the strings 
ending on BPS branes by now seems to be well understood in the flat worldvolume case, 
the problem of formulating and solving it for non flat branes is still left to be 
completely settled out.
A first major step in this direction has been done in \cite{vafa}
where the naturalness of the twist mechanism for the gauge theory 
living on the brane was noticed. The root of this stems to the fact that
the transverse bosonic degrees of freedom are in fact sections of the normal 
bundle of the embedded hypersurface on which the branes lay down.
On the other hand, the BPS stability condition is guaranteed 
only if on the brane worldvolume some supersymmetry is left \cite{bbs}.
In \cite{vafa} the supersymmetric cycle condition 
has been indicated as the root of the twisting procedure.

All this finds a natural explanation once linked to the geometrization
of the effective field theory point of view introduced in \cite{mto}
where the realization of the world-volume embedding equations has been 
understood as the realization of the SW curve associated to the relevant 
gauge theory.
SW curves \cite{SW} have been introduced as basic tools to understand 
the geometry of the space of gauge inequivalent vacua in supersymmetric 
gauge theories. Their RG properties and phases have been therefore better 
understood in a proper geometric framework. 
An important development \cite{DW} has been given 
by the understanding that these curves can be seen as spectral curves of 
an integrable system and that the SW differential generates the RGEs
as an explicitly integrable system.
On the other hand these curves constitute the materialization of 
flat D-branes world volumes in the M-theory flat target space.
As far as we are concerned, the most important property of this
construction is that the curves intrinsic stability has been understood in 
terms of the stability of the corresponding vacuum in the gauge theory.
This means that the moduli space of these 
curves represent stable gauge inequivalent vacua of the theory -- at 
least in the YM-coupling regime which we believe to be covered by this picture.
Let us remark that
all this just follows from the fact that the associated 
integrable system is obtained by the equations 
associated to susy preserving BPS field configurations. 

Due to the uncontestably partial view offered by perturbative formulation of 
string theories, 
during the last years the interest has been extended to the study of its non 
perturbative
sectors. The concept of strong-weak duality led to a new picture of the 
string theory
moduli space which goes under the name of M-theory \cite{Mth}.
The spectrum of the D=11 corner of M-theory contains membranes and M5-branes states
whose dynamics has not been fully understood by now.
Under compactifications M-theory generates, in particular, all D-brane states.
One might be interested in understanding if the integrable system framework
can then be promoted in some way also to the study of M-branes states.

A really odd object in M-theory is represented by the 
M5-brane
since, due to the self-duality of the relevant 2-tensor which lives on it, 
its relative gauge theory is of a non lagrangean type and then much harder 
to be 
understood. On top of it, another and harder problem seems to be raised by 
the lack of a proper  
formulation in the case of (non-abelian) higher rank gauge theories which should apply 
to the multi M5-brane theory.

As far as the single M5-brane theory is concerned, the determination of 
its partition 
function is available, at least in some specific cases, 
but its multibrane analogous still resists several attaches.

A simplified configuration, which is more suitable to be studied, is the 
case in which the M5-brane world-volume is of the factorized form $T^2\times M_4$,
where $T^2$ is a two-torus and $M_4$ a four-manifold.
In this case \cite{D1} 
the analysis of the CFT living on the M5-brane can be 
attached by reducing to the limit of vanishing two-torus volume and landing
on a four dimensional gauge theory.

In the case which we will study,
the resulting gauge theory is the twisted version of SYM
with ${\cal N}=4$ and $U(N)$ gauge group which has been 
considered in \cite{vw}.
Let us notice here that once the spectrum of the theory is given, its two 
derivative low energy effective theory comes out to be automatically topological.
This suggests that in such compactifications (see later on for a more precise
picture) there is a substantial decoupling of the worldvolume theory from the bulk 
parameters so that in makes sense to consider the N M5-branes bound state in isolation.
 This is possible if in particular the corresponding M5-brane anomalies 
vanish.

On the other hand the same theory can be studied in an equivalent limit of
vanishing $M_4$ volume thus showing the naturalness of the results in \cite{vw}
where the structure of the partition function was exhibited as corresponding to 
a two dimensional toric model \cite{V}.

In this paper we will try to add a new piece to this brane -- gauge theory
correspondence by proposing a solution for the twisted ${\cal N}=4$ 
$U(N)$ theory with $M_4$ a simply connected Kaehler four manifold.
Our analysis generalizes the one given in \cite{vafa'} 
where $M_4$ was a $K3$ surface, to the case in which $M_4$ is a simply connected
Kaehler manifold with $h^{(2,0)}>0$. 
This will be done by exploiting the structure of the gauge inequivalent 
vacua space 
of the theory as a space of holomorphic covering of the base four 
manifold.

This article is organized as follows.
In the next section we will recall how the twisting gets generated in generic CY case 
and exploit the natural link between twisted Higgs-Hermitian configurations
describing D-brane bound states and spectral covers.
In section 3 the elliptic genus associated to a single M-5brane is discussed in full detail
in order to ideologically justify the subsequent developments.
In section 4 the case of N M5-branes is solved by the explicit solution of the relevant 
moduli space identification problem.
Conclusions and open questions are contained in section 5.

\section{Refreshing the twist and the spectral covers}

Let $M_D$ be a D dimensional manifold (D=10 for type II theories and D=11
for M-theory) and
let $W_p$ be a $p+1$ dimensional hypersurface smoothly embedded in $M_D$
with $p+1\leq D$.

If $W_p$ is flat, the low energy effective theory of N Dp-branes in type II 
theories whose worldvolume coincides with $W_p$ is the p+1 dimensional 
reduction of the ${\cal N}=1$ SYM in 10 dimensions.
This represents, at weak coupling, the theory of open strings ending 
on the branes. In particular
the transverse D-p-1 bosonic fields, which are
in the adjoint of the gauge bundle and vectors with respect to the 
transverse unbroken rotational group, are interpreted as representing the 
transverse motions of the brane itself.

Let us now turn to the more complicate case in which $W_p$ is 
not a flat manifold. In particular it means that there is no possible choice
for the embedding functions to produce a flat induced metric.
It is possible that the normal bundle is nonetheless flat, due to some 
particular structure of the embedding. In this case one is led to study
SYM theory on the curved $W_p$ and then the analysis carried out in
\cite{blau} applies. We will concentrate here in cases in which 
the normal bundle is not flat.

The generic situation arises as follows. 
Let us consider a given $M_D$ target space preserving some fraction of susy.
This implies the existence of covariantly constant spinors on it
and hence an holonomy reduced structure in the target space.
On the other side, the susy left on the brane sitting on $W_p$
will be determined by the relative orientation of the 
reduced spin bundle under the reduced holonomy. 
The supersymmetric cycle condition \cite{bbs} then 
relates the normal and the tangent bundles
in a non trivial way. This implies that the spinor bundle
${\cal S}=ST(W_p)\otimes SN(W_p)$, to which the spinors belong, admits
a representation as a sum of integer spin representations of the 
tangent bundle itself. This induces then the twisting 
of the supersymmetric SYM theory which would be present in the flat case. 
In particular
the superalgebra itself get twisted to the appropriate 
cohomological algebra.
Several examples of the above mechanism can be found both in
\cite{vafa} and in \cite{bt}, where a general discussion about the induction of 
the twist by the normal bundle is presented.

Let us concentrate on (partial) wrapping along even susy Kaehler cycles. 
Notice that, while for a single brane the BPS stable states 
are specified by particular holomorphic embeddings in the target space,
in the $N$ branes case the BPS stable states are determined by
an integrable system. In several cases the normal bundle splits as
$N(W_p)={\cal N}(W_p)\oplus\bar{\cal N}(W_p)$ and the integrable system
is built from an holomorphic vector bundle $E$
and an holomorphic section $\Phi^{\cal N}$ of $Adj(E)\otimes {\cal N}(W_p)$
as
$$
{\cal D} \Phi^{\cal N}=0
\, ,\quad
F^{(2,0)}=0\, ,\quad \omega\cdot F+
\left[\Phi^{\cal N},\left(\Phi^{\cal N}\right)^\dagger\right]=0
\, ,$$
where $\omega$ is the Kaehler form on the cycle.

In particular then the spectral cover for the system
$$
{\rm Det}\left(\Phi^{\cal N}-\phi{\bf 1}_N
\right)=0
$$
materializes in the total space of the 
holomorphic factor of the normal bundle and classify its solutions.
Let us here stress the obvious fact that the normal bundle total space is nothing but 
(a part of) the target space itself. Hence
this curve is identified with the relevant wrapped worldvolume
of the N branes system and is the analogous of the SW curve in the 
curved case.

The case of the M5-brane is technically more complicated and much less understood, but we accept
it to follow a similar pattern. 
The first problem is that the worldvolume theory
is not a SYM theory, but a gauge theory of a self-dual tensor multiplet.
Secondly, the higher rank case is not either formulated.
In this case, one can try to understand it as the strong coupling limit 
of the D4-brane theory, but the very structure of the M-interaction 
\cite{Hull} remains unknown by now.
A possibility to investigate the M5-brane theory structure is anyhow to 
wrap it on a product cycle in such a way to produce a twisted supersymmetric 
YM theory, which is then topological, on a factor. 
Assuming it not depending on the ratio of the volumes of 
the factor cycles, it is then possible to try to gain some informations just by 
rescaling inversely the volumes and see if the theory happens to have a simpler picture.
Notice that so doing we are pretending the 
world-volume theory of the M5-brane to be effectively decoupled 
with respect to the bulk fields. On the other hand it is well known that 
the theory of the M5-brane suffers from inborn and inflow anomalies
and that therefore, to be have a consistent substantially decoupled
brane configuration one has to verify  that both these contributions to 
the anomaly cancel separately.
Of course, the coherence of the full out-come result is a test for the 
M-theory conjecture.

\section{The single M5-brane partition function}

The geometric set-up that we refer to is the following \cite{msw}.
We consider M-theory on $W=Y_6\times T^2\times R^3$, where
$Y_6$ is a Calabi-Yau threefold of general holonomy.
Let $M_4$ be a supersymmetric simply connected four-cycle in $Y_6$
which we take to be a representative of a very ample divisor \cite{msw}.
Notice that $M_4$ is automatically equipped with a Kaehler form
$\omega$ induced from $Y_6$.
We consider then one 5-brane wrapped around $C=T^2\times M_4$.

The bosonic spectrum of the world-volume theory of this 5-brane is given by
a 2-form $V$ with self-dual curvature and five real bosons taking values
in the normal bundle $N_C$ induced by the structure of the embedding
as $T_W|_C=T_C\oplus N_C$. Passing to the holomorphic part and to 
the determinants and using the properties of $Y_6$,
it follows that the five transverse bosons are respectively, three non-compact 
real scalars $\phi_i$
and one complex section of $K_{M_4}=\Lambda^{-2} T^{(1,0)}_{M_4}$,
the canonical line bundle of $M_4$, $\Phi$.
The (partially) twisted chiral $(0,2)$ supersymmetry completes the spectrum.

Notice that to have a self-consistent theory of a 5-brane in isolation,
it has to be anomaly free by itself and also from the inflow point of view.
These conditions are fulfilled if\cite{anomalie}
$$\left(p_1(T_C)-p_1(N_C)\right)^2=4p_2(T_C)\quad {\rm and}
\quad p_2(N_C)=0.$$
In our case we have $$p(T_C)=c(T^{(1,0)}_{M_4})c(T^{(0,1)}_{M_4})=1+(2c_2-c_1^2)
+c_2^2$$
$$p(N_C)=c(K_{M_4})c(\bar K_{M_4})
=1-c_1^2$$ 
where $c_i=c_i\left(T^{(1,0)}_{M_4}\right)$,
and the above conditions are identically satisfied. Let us notice 
here that this requirement is enough also for the cancellation of the 
anomalies in the case of several 5-branes, at least as it is given in \cite{Nanomaly}. 

Even if a precise recipe to give a lagrangean formulation of the theory 
is not known, fortunately we have the possibility to calculate
the partition function of the 5-brane in the limits in which the volume of 
one the two factors of $C=T^2\times M_4$ vanishes. If the two results will 
agree we can believe that really the theory depends on the product
of the two volumes only and then promote this feature to the $N>1$ case
too.

Let us reduce to zero the volume of $M_4$. In this case
\cite{V,G} $V$ gives a vector $a$, $b_+$ left
and $b_-$ right real compact scalars, $\Phi$ gives $b^{(0,2)}$
complex scalars and $\phi_i$ three real scalars.
All in all we have $b_2+2$ left and $4b^{(0,2)}+4$ right scalar bosons.
The chiral fermions will provide correspondingly $4b^{(0,2)}+4$ right 
periodic fermions and no left fermions.

The elliptic genus is defined as
$${\cal E}=
\frac{(Im\tau)^{d/2}}{V_d} {\rm Tr}_{RR}
\left[
(-1)^FF_R^{\sigma/2}q^{L_0}\bar q^{\bar L_0}
\right]\, ,
$$
where $d$ is the number of non-compact scalar bosons, $V_d$ their zero-mode volume and 
$(0,\sigma)$
are the supersymmetries of the model. By general arguments, ${\cal E}$ is a 
$(-d/2,-d/2)+(0,\sigma/2)$ modular form.

In our case we have $\sigma=2b_++2=4b^{(2,0)}+4$
right fermions and $d=3+2b^{(2,0)}=2+b_+$ non-compact scalar bosons
by the dimensional reduction giving the elliptic genus to be a modular form 
of total weight $(-1-b_+/2,b_+/2)$.

We can calculate the elliptic genus using standard tecniques in $CFT_2$ 
as\footnote{We assumed $b_1=0$ because of the ampleness of $M_4$ as a divisor in $Y_6$.
If this would not be the case, we then had a
further multiplicative factor $\left(\eta/\bar\eta\right)^{2b_1}$.}
$$
Z_1^M=
\left(\frac{\theta_\Lambda(q,\bar q)}{\eta^{b_-}(q)\eta^{b_+}(\bar q)}\right)
\cdot
\left(\frac{1}{\eta(q)\eta(\bar q)}\right)^{3+2b_{(2,0)}}
\cdot
\left(\eta(\bar q)\right)^{4b_{(2,0)}+4} 
=\frac{\theta_\Lambda(q,\bar q)}{\eta^\chi(q)}
$$
where the first factor comes from the compact bosons, the second from the non compact 
bosons and the last one comes from the fermions. The function
$$
\theta_\Lambda(q,\bar q)=
\sum_{m\in\Lambda}q^{\frac{1}{4}(m,*m-m)}{\bar q}^{\frac{1}{4}(m,*m+m)}
$$
is a generalized $\theta$-function,
$\Lambda$ is the lattice of integer period in $H^2(M,R)$, $(\cdot,\cdot)$
is the intersection form, $*{}m$ is the Hodge dual of $m$ and 
$q=e^{2\pi i\tau}$.
The unmatched left scalar bosons produce the overall
factor of $\eta(q)^{-\chi}$, where
$\eta(q)=\sum_n(-1)^nq^{\frac{3}{2}(n-\frac{1}{6})^2}$ is the Dedekind 
$\eta$-function and
$\chi=2+b_2$ is the Euler characteristic of $M$.
In particular $Z_1^M$ is a modular form of weight $(b_-/2,b_+/2)+(-\chi/2,0)=
(-1-b_+/2,b_+/2)$ as it had to be.

Let us now reduce to zero the volume of the torus.
The six-dimensional $V$ field reduces then to a vector $A$ and to a real 
scalar $b$,
while $\Phi$ just generates a section of the canonical bundle
$B$ and the three scalars $\phi_i$ remain three scalars.
The twisted $U(1)$ theory spectrum is given by the gauge field $A$, 
a self-dual 2-form $B^+=B+\omega b+\bar B$ and three scalars, 
as far as the bosonic part is concerned, and by
its twisted-susy counterpart given by Grassmann valued two self-dual 2-forms, 
two 1-forms and two scalars. 

Up to the gauge field $A$, which is a connection for the gauge bundle 
all the other fields are also sections of the adjoint bundle, and therefore 
in the single brane case, uncharged.

As it is naturally expected, in the calculation of the $U(1)$ partition 
function, once also the ghosts are properly taken into account, there is 
a complete cancellation between all the fluctuation contributions 
and we are left with an exact expression of the partition function 
coming only from the zero-mode contributions
\footnote{In details, 
denoting by $\Delta_i$ the determinants of the Laplacian on $i-$forms,
the bosonic sector contributes with
$\frac{\Delta_0}{\sqrt{\Delta_1}}\cdot\frac{1}{\sqrt{\Delta_2^+}}\cdot
\frac{1}{(\Delta_0)^{3/2}}$, while the fermionic sector contributes
as $\left[(\Delta_2^+)^2\cdot(\Delta_1)^2(\Delta_0)^2\right]^{1/4}$
giving all in all $1$.
Also an exact cancellation between the zero modes volumes takes place
and no ${\rm Im}\tau$
factors are present because of the topological symmetry of the twisted theory.}.
Therefore, the contribution from the gauge sector is the classical 
one coming from the $U(1)$ gauge field which can be evaluated to be \cite{W}
$$
q^{-\chi/24}\theta_\Lambda
$$ where
$
\theta_\Lambda(q,\bar q)=
\sum_{m\in\Lambda}q^{\frac{1}{4}(m,*m-m)}{\bar q}^{\frac{1}{4}(m,*m+m)}
$
is exactly the same object that we have found before \cite{V}.

To obtain the full partition function one has to add up 
also the contribution coming from 
point-like degenerate instantons which amounts to a multiplicative factor of
$$\sum_n q^n {\rm dim}H^*(M_4^n/S_n)=\frac{1}{\prod_{n>0}(1-q^n)^{2+b_2}}
=q^{\chi/24}\eta^{-\chi}\,.$$
The nature of this factor can be traced back to \cite{ns}.

Therefore, all in all, we obtain
$$
Z_1^M=\frac{\theta_\Lambda}{\eta^\chi}\,
$$
which is the same result that we had in the two dimensional computation.

\section{The $N$ M5-branes partition function}

The N 5-brane theory is not known. What we assume to be true -- and will show to be 
consistent -- is the following.

By shrinking the $M_4$ volume to zero we find a toroidal $\sigma$-model
valued in the moduli space ${\cal M}$ of BPS-like field configurations of the twisted
${\cal N}=4$ SYM $U(N)$ theory on $M_4$ itself, in the spirit of \cite{BJSV}.
By shrinking the torus volume to zero we find directly the twisted
${\cal N}=4$ SYM $U(N)$ theory on $M_4$.

The torus contribution will amount to the elliptic genus of the (stratified) 
moduli space ${\cal M}$, while the same object is to be obtained as the
$M_4$ partition function (with fermionic insertions). This, as explained 
in \cite{vw}, counts the Euler characteristic of ${\cal M}$ and fits
exactly this identification \cite{D1}.

Notice that in the limit of zero volume of the torus, M-theory on $T^2\times Y_6\times R^3$
is dual to perturbative type IIB on $R\times Y_6\times R^3$ with coupling $\tau$ and that
the N M5-brane system gets mapped to a bound state of N D3-branes wrapped around $M_4$.
Their low energy description is in fact given by the twisted ${\cal N}=4$ SYM with
gauge group $U(N)$ as given above.

\subsection{The moduli space and the spectral cover}

In this section we calculate the relevant moduli space of the 
twisted ${\cal N}=4$ Yang-Mills theory. 

The twisted $U(N)$ theory spectrum is given by the gauge field $A$, a self-dual 
2-form $B^+$ and three scalars, as far as the bosonic part is concerned, and by
its twisted-susy counterpart given by Grassmann valued two 2-forms, two 1-forms
and two scalars. Beside the gauge field $A$, which is a connection for 
the gauge bundle 
$E$, all other fields are also sections of the adjoint bundle $Adj(E)$.

Let us start from the susy fixed point condition for the twisted 
$\N=4$ theory with gauge group $U(N)$. 
These are given by the following equations \cite{vw}
\be
\omega\wedge F +[B,\bar B]=0
\quad
F^{(2,0)}=0
\quad
\bar D B=0
\la{1}\ee
where $\omega$ is the Kaehler form on $M$, $F$ is the curvature
of the relevant gauge bundle $E$, $B$ is a section of 
$\End(E)\otimes K$ with $K$ the canonical line bundle of $M$
while $\bar B$ is its conjugate.

We associate to each solution of (\ref{1}) a spectral four manifold 
defined by the equation
\be
\det(k\1-B)=0\, ,
\la{2}\ee
where $k$ is an indeterminate taking values in the total space of the 
canonical line bundle $K$ where eq.(\ref{2})
defines a rank $N$ holomorphic covering of $M$.
Let us call $\Sigma$ this resulting Kaehler manifold.

To construct the solution explicitly, start writing the spectral surface 
equation as 
\be
0=\det(k\1-B)=k^N+\sum_{i=0}^{N-1} k^i\beta_i
\la{4}\ee
where $\beta_i$ are holomorphic sections of $K^{N-i}$ which
can be explicitly be given by symmetric polynomials in $B$.

Let us first consider the case in which $\Sigma$ is connected.
This means that the polynomial in (\ref{4}) is irreducible.
By using the complexification of the gauge group, we can reduce 
$B$ to its bare bones, i.e. write it as 
\be
B=
\left(\matrix{
-\beta_{N-1}&-\beta_{N-2}& \dots &\dots          & -\beta_0\cr
1           & 0          & \dots &\dots          & 0       \cr
0           & 1          & 0     &\dots          & 0       \cr
:           & :          & :     & :             & :       \cr
0           & 0          & \dots & 1             & 0       \cr
}
\right)
\la{5}\ee
while the reduced complex bundle remains
\be
E=E_N=\oplus_{i= 1}^N L^{N+1-2i}
\la{6}\ee
where $L$ is a line bundle over $M$ such that $L^2=K$.
This data reduction from principal Higgs bundles spectral surface data
is called abelianization in the mathematical literature (see for example
\cite{sc}).
Notice that only if $N$ is even $K$ has really to admit a square root $L$.
Moreover, the overall phase is fixed by the $SL(N,{\bf C})$ complexified 
structure group. 
As we will explain later on, it is not at all a coincidence that
$E_N$ in (\ref{6}) is modeled on the $N$ dimensional irreducible $SU(2)$
representation.

In general the $B$ characteristic polynomial splits in irreducible 
monic factors\footnote{For example, let us consider the case $N=2$
and the polynomial $k^2+b_0$ where $b_0$ is a holomorphic section of $K^2$.
We have factorization if we can write $b_0=s^2$ with $s$ a holomorphic section of $K$.}
of degree $a>0$ as
\be
\det(k\1-B)=\prod_a [P_a(k)]^{n_a}
\la{10}\ee
where $N=\sum_a a\cdot n_a$.
The relative covering spectral surface factors then in connected 
components as $\Sigma=\cup_a [\Sigma_a]^{n_a}$, where
$\Sigma_a$ is the surface given by the equation $P_a=0$ and is given by a
specific choice of $L$.

Let us notice that the multiplicities $n_a$ indicates the presence of a 
non trivial unbroken subgroup 
$J=S_{n_1}\times\left(S_{n_2}\bowtie{\bf Z}_2\right)
\times\left(S_{n_3}\bowtie{\bf Z}_3\right)
\dots$ of the 
$U(N)$ Weil group $S_N$ which has to be factorized out in order to get the 
true moduli space.
This determines the moduli space of solutions of (\ref{1}) to be the Hilbert scheme of 
holomorphic coverings of $M_4$ in $Y_6$.

The factorization (\ref{10}) induces the analogous 
structure on the vector bundle as 
$E=\oplus_a E_a^{\otimes n_a}$ which is
in fact modeled on the generic reducible 
N dimensional representation $R_N$ of $SU(2)$. 
In terms of irreducible ones we have in fact
$R_N=\sum_a [R^{(irr)}_a]^{n_a}$, where $R^{(irr)}_a$ is the unique $a$
dimensional $SU(2)$-representation. The sum over these $SU(2)$ embeddings,
which will not be reviewed here,
is the sum over the sectors explained at length in \cite{vw} (section 5)
and in \cite{vafa'}.
Let us point out here that this analysis \cite{w4} of the vacua structure
fails exactly at the canonical class, which for a generic Kaehler
manifold is non empty, giving rise to locally enhanced gauge symmetry.
This corresponds exactly in the spectral cover picture to the branching locus.
Notice also the natural action of the $J$ subgroup of the Weil group on $R_N$
which symmetrizes with respect to the multiplicities of the $SU(2)$ irreps.
Anyhow, this field theory analysis is valid only if $b_+^{M_4}>1$ when the moduli space of 
holomorphic deformations of $M_4$ in $Y_6$ is of positive dimension.

Generically the canonical divisor $K=\sum_l C_l$ can be considered to be a sum of 
irreducible curves $C_l$. The branching locus
of the covering is the zero locus of 
the discriminant of the B characteristic polynomial (\ref{2}) which is a 
holomorphic section of $K^{N(N-1)}$. Therefore the branching locus can be represented 
as a collection of the irreducible components of the canonical divisor itself as 
$H=\cup_l' C_l'$ for a subset $\{l'\}\in\{l\}$.
Unfortunately we cannot still conclude about the structure of the
spectral cover canonical class, $K_\Sigma$ because of a residual set in the lifting
of the branching locus. On the other hand, because of ampleness,
we have in any case that 
$K_\Sigma$ is still very ample by construction and 
therefore its Betti numbers are computable directly by making use of the formulas 
obtained in \cite{msw} as
$$
b_+(\Sigma)=-1+\int_Y \frac{1}{3}\Sigma^3+\frac{1}{6}\Sigma c_2(Y)
$$ $$
b_-(\Sigma)=-1+\int_Y \frac{2}{3}\Sigma^3+\frac{5}{6}\Sigma c_2(Y)
$$
and the fact that $\Sigma=M^N/H$.
Specifying $Y$, $M$ and $H$ one can in principle explicitly calculate also the
intersection form on $\Sigma$.

\subsection{$K3$ as the totally reducible case and the $T^2$ multiplicity}

Let us now go back to the 5-brane point of view:
the 5-brane is taken to be $N$ times wrapped around $W=T^2\times M_4$.

In \cite{vafa} it has been shown that if $M_4$ is a K3, 
then the partition function of $N$ coinciding branes on it
is obtained by summing over all the possible rank $N$ 
holomorphic coverings of the $T^2$ over itself.

If $M$ is a $K3$ surface, then, since $K3$ does not admit any non trivial 
holomorphic line bundle for a generic complex structure, 
there is no non trivial solution of (\ref{1}).
Therefore, in calculating the N 5-branes partition function,
the sum over all the possible holomorphic coverings of $C$ over itself 
reduces to the $T^2$ sector since the only irreducible holomorphic covering of K3
results to be the trivial one.
The counting is then made following \cite{dmvv} by summing over all the possible 
holomorphic coverings of $T^2$ over itself, which is by summing over all the resulting 
values of the covering tori moduli, the result being given by the Hecke
transform of order $N$ of the single 5-brane partition function
on $W$
\be
Z_N^{K3}=\H_N Z_1^{K3}
\la{3}\ee
Let us comment about the important fact that the M5-brane point of view 
tells us the exact way we have to count the multiplicity of the 
trivial covering, i.e. it is the M5-brane point of view which fixes the form of the 
Hecke transform induced by the symmetric space elliptic genus formula.

Our aim is now to generalize the above result to a more general case.
The counting of BPS bound states of N M-5branes is then 
the counting of holomorphic self-covering of $C$
on itself with total rank $N$. They can be classified
in terms of partitions of $N$ in positive integers as $N=\sum_a a\cdot n_a$ 
such that each term in the sum represents a connected component of the 
covering and 
each component is a connected $n_a$-folded holomorphic covering of the torus
times a connected $a$-folded holomorphic covering of $M_4$.

Let us now associate to each connected component a partition function $Z_{a,n_a}$.
Then, for each given partition $N=\sum_a a\cdot n_a$, 
the total contribution to the partition function 
will be given by the product
$$
\prod_aZ_{a,n_a}\,.
$$
From the $K3$ case analysis we learn that
the $T^2$ multiplicity $n_a$
is naturally kept into account by the Hecke 
operator as
$Z_{a,n_a}=H_{n_a}Z_{a,1}$, where $H_n$ is the Hecke transform of order $n$.
The Hecke operator structure can be recognized from the field theory side
as being the result of the process of redefinition of the coupling of the gauge
theory due to the rescaling of the traces weights \cite{vafa}
and we have shown it to correspond just to the $J$ symmetrization operation.
On the other side, its appearance from the torus point of view is naturally obtained as 
the result of the relevant instanton sum in the $\sigma$-model \cite{bcov}.
Let us anticipate here that the contribution $Z_{a,1}$ will turn out to be a modular 
form of weight $(-1-b_+^{\Sigma_a}/2,b_+^{\Sigma_a}/2)$.
Since the Hecke transform preserves modularity, our final result admits a natural
interpretation from the two dimensional point of view as the elliptic genus 
of a specific sigma model valued in the coomology of the moduli space
of vacua that we just calculated. The calculation is much similar to the
one that we already gave for the rank one case.

We are then left with the calculation of the four dimensional field theory
contribution in the irreducible case without torus multiplicity.

\subsection{The field theory counting}

By the topological nature of the twisted ${\cal N}=4$ SYM, our
partition functions localize around the solutions of (\ref{1}).
Therefore, if one is going to calculate the partition function of 
$N$ coinciding 5-branes wrapped around $C$ as the partition function of 
a single 5-brane wrapped $N$ times around $C$ by counting all the 
possible rank $N$ holomorphic coverings of $C$ on itself,
then the relative spectral surface should be kept into 
account in the covering counting, if solutions with non trivial $B$ 
of (\ref{1}) exist.

Let us now proceed by making use of 
the analysis of the vacua structure of the theory that we just performed.
As we have shown above all the leftover calculation is reduced to
the irreducible coverings.

The field theory path integral for the twisted susy theory can be performed 
by semiclassical approximation which, due to the topological features 
of the twisted theory,
turns out to be exact. This can be done just following the analysis in \cite{bbtt},
which is given for the untwisted case, adapting it to the twisted $U(N)$ theory
as follows.
Each generic point in the (irreducible part of the) moduli space 
${\cal M}_{irr.}$ breaks completely the gauge invariance from $U(N)$
to the maximal compact Cartan torus $U(1)^N$, up to the local enhancing at the 
canonical class, by singling out a canonical choice for a Cartan subalgebra.
Integrating out the non-Cartan valued fields, the semi-classical
evaluation of the path integral reduces to a collection of $N$-abelian
twisted multiplets, one for each sheet of the spectral cover.
The lifting technique, which is just the field theory counterpart of the 
abelianization for sections of principal Higgs bundles, then gives as an outcome
the twisted super-Maxwell theory on the spectral cover $\Sigma$,
which is the single 5-brane contribution $Z_1^\Sigma$ that we just had in the previous 
section, but
evaluated on the spectral cover\footnote{The lifting technique has been originally 
performed in the two dimensional case corresponding to Matrix String Theory
in \cite{MST}.}.

The total contribution to the irreducible sector
will be then given after a summation over all the possible $L$ line bundles is 
also kept into account.
Assuming $K^{1/2}$ to exist on $M_4$, we have $L=K^{1/2}\otimes {\cal O}_\varepsilon$.
Here ${\cal O}_\varepsilon=\sum_l \epsilon_l [C_l]$, where $[C_l]$ are the reductions 
modulo 2 of (the Poincare' dual of) the curves $C_l$ that we introduced before,
the label $\varepsilon$ indicates the collections of integer
numbers $\{\epsilon_l\}$ which satisfy $\epsilon_l^2=\epsilon_l$ for each $l$.
Recalling that the reduced form of the vector bundle is
$E_N=\sum_{i=1}^N L^{N+1-2i}$, we obtain $E_N={\cal O}_\varepsilon^{\otimes N+1}\otimes
\sum_{i=1}^N (K^{1/2})^{N+1-2i}$.
With this parametrization of the line bundle $L$ we can write our final result as
$$Z_{a,n_a}={\cal H}_{n_a}Z_{a,1}\quad {\rm and}\quad Z_{a,1}= 
\sum_{\varepsilon}\frac{\theta_{\Lambda^{\Sigma_a}+x}}{\eta^{\chi_{\Sigma_a}}}$$
where $x=[{\cal O}_\varepsilon^{\otimes a+1}]$ shifts correspondingly the
lattice of integer periods $\Lambda^{\Sigma_a}$ on $H^2(\Sigma_a,R)$. 
Notice that this shift is effective only if $a$ is even.

Needless to say, $Z_{a,1}$ is non vanishing effectively
only if there exists irreducible solutions of the spectral equation
of the corresponding orders.

A closer inspection should reveal the correspondence between the sum over the
$x$-classes, once the proper $U(1)$ components are divided out from our formula, 
and the sum over the SW classes as obtained in \cite{D2} for the $SU(2)$ case.
Moreover, similar results to the
$K3$ case should hold more generally for manifolds $M$ and given values of $N$ 
such that $M$ does not admit irreducible self 
holomorphic coverings of ranks corresponding to partitions of the actual
$N$, beside the one corresponding to $a=1$ and $n_1=N$.

Since the partition function $Z_1^{\Sigma}$ is completely determined by 
the intersection form on $H^2(\Sigma)$, we could calculate it in full explicit form
case by case by specifying the particular cycle in a given Calabi-Yau.

\section{Conclusions and open problems}

In this note we used the geometrical properties of 
the M5-brane in M-theory to address the problem of calculate
the partition function in the twisted ${\cal N}=4$ D=4 SYM
on a Kaehler manifold, pushing on a strong geometric interpretation 
of the first.

Our calculational scheme is quite general and could be tested in different 
specific configurations. In particular, we think it would be important 
to understand weather
it extends to the $b_+^M=1$ cases where the absolute rigidity
of the 4-cycle causes well known problems in the evaluation of the partition 
function, the quantum field theory vacua analysis remaining empty.

It would be very interesting to understand from this general point of view 
also the problem of holomorphic anomalies.
This question is strictly linked to the degeneration corners of the moduli 
spaces in the reducible locus and can be understood from the spectral cover point
of view almost easily in terms of subloci in which the factorization in different 
connected components of the generically connected spectral surface takes place.
This automatically generates extra zero-modes representing other M5-branes 
bound states at threshold causing the holomorphic anomaly.
The technical and less trivial part is then the full identification of 
that contribution in the partition function.

The most severe bound in performing the above programs is that admittedly
our formula is rather implicit because of the lack of the necessary 
mathematical technology to calculate in the generic case the canonical class 
of the spectral surface and its intersection form.

Another interesting possible development could be to try to extend 
this analysis to M-theory on $R^3\times Y_8$, with $Y_8$ an elliptically fibered 
CY 4-fold and the M5-branes wrapped around an elliptically fibered 
supersymmetric 6-cycle in it in order to try to extract more informations about the 
structure of the N M5-brane world-volume theory.

With the same attitude, a possible very interesting development would be to
generalize and clarify the genus expansion given for the $SU(2)$ case in
\cite{D2} to the $U(N)$ case obtained here. Doing this, one could be able to 
exhibit the form of the N M5-brane partition function as a (tensionless) string
theory one.

As a possible application one could use the results obtained here to M-theory 
black hole entropy counting generalizing the analysis in \cite{msw,vafabh}.
The result of our investigation suggests the existence of several contributions 
relative to the different possible holomorphic covering of $C=T^2\times M_4$
in the form of a fine structure in the entropy formula.
This seems to lead to subleading M-theory corrections to the full 
eleven dimensional supergravity result as discussed in \cite{ms}.
Anyhow, all the sectors counting, done as
$$
(\Delta S)_{a,n_a}=2\pi\sqrt{\frac{n_a c_L^{\Sigma_a}}{6}}\, ,
$$
where $c_L^{\Sigma_a}=\int_{Y_6}(\Sigma_a^3+\Sigma_a\cdot c_2(Y_6))$,
will yield, in the large cycle volume and large $N$ limit, the appropriate result.

\vspace{1 cm}
{\bf Acknowlodgments}
I would like to thank M.~Bertolini, L.~Bonora, A.M.~Boyarsky, R.~Dijkgraaf, 
A.~Hammou, E.~Looijenga,
J.F.~Morales, A.~Tomasiello and M.~Trigiante for interesting discussions.

Work supported by the European Commission RTN programme
HPRN-CT-2000-00131.

\end{document}